\documentclass[11pt]{article} 
\usepackage{hyperref} 
\pdfoutput=1
\begin{document}
 
\title{Bubble Dynamics in a Two-Phase Medium}

 \author{ Arvind Jayaprakash, Sowmitra Singh, and Georges Chahine  \\
\\ DYNAFLOW, INC.,\\ www.dynaflow-inc.com \\
10621-J Iron Bridge Rd,  Jessup MD 20794 \\
Email: arvind@dynaflow-inc.com 
}
 \maketitle

\begin{abstract} 

The spherical dynamics of a bubble in a compressible liquid has been studied extensively since the early work of Gilmore.  Numerical codes to study the behavior, including when large non-spherical deformations are involved, have since been developed and have been shown to be accurate.  The situation is however different and common knowledge less advanced when the compressibility of the medium surrounding the bubble is provided mainly by the presence of a bubbly mixture.  This applies to both when the studied bubble is just one of the bubbles composing the two-phase medium, or in the simpler case where the concerned bubble is much larger than the finer bubbles forming the surrounding medium and is the driving force to the dynamics of this medium.  In both cases, modeler resort to simplifying assumptions, with need for validation, would benefit from a basic fundamental study combining experimental observations and numerical/analytical simulations.  This is more so since almost all simulation models use some form of single bubble dynamics as the building block to the two-phase models. 

In one of the present works being carried out at DYNAFLOW, INC., the dynamics of a primary relatively large bubble in a water mixture including very fine bubbles is being investigated experimentally and the results are being provided to several parallel on-going analytical and numerical approaches. The main/primary bubble is produced by an underwater spark discharge from two concentric electrodes placed in the bubbly medium, which is generated using electrolysis. A grid of thin perpendicular wires is used to generate bubble distributions of varying intensities.  The size of the main bubble is controlled by the discharge voltage, the capacitors size, and the pressure imposed in the container.   The size and concentration of the fine bubbles can be controlled by the electrolysis voltage, the length, diameter, and type of the wires, and also by the pressure imposed in the container.  This enables parametric study of the factors controlling the dynamics of the primary bubble and development of relationships between the bubble characteristic quantities such as maximum bubble radius and bubble period and the characteristics of the surrounding two-phase medium: micro bubble sizes and void fraction.  The dynamics of the main bubble and the mixture is observed using high speed video photography.  The void fraction/density of the bubbly mixture in the fluid domain is measured as a function of time and space using image analysis of the high speed movies. The interaction between the primary bubble and the cloud are analyzed using both field pressure measurements and high-speed videography. Parameters such as the primary bubble energy and the bubble mixture density (void fraction) are varied, and their effects studied. 

The attached fluid dynamics \href{run:.\\anc\\Dynaflow-Bubble_in_Bubbly_Media-HI.mpg}{Video}  shows the dynamics of a primary spark bubble expanding and collapsing with/without a bubbly medium. The ambient pressure is set to 174mBar and the strength of the spark used is 11kV.  It has been observed, both numerically and experimentally, that the presence of small bubbles dampens the growth of the central/primary bubble besides decreasing its collapse time [1, 2]. 

\begin{enumerate} 

\item

 Jayaprakash, A., Singh, S., and Chahine, G. L., "Bubble Dynamics in a Two-Phase Bubbly Mixture", IMECE2010-40509, ASME 2010 International Mechanical  Engineering Congress \& Exposition, Vancouver, British Columbia, November 10-12. , 2010
 
\item 

Raju, R., Singh, S., Hsiao, C.-T., and Chahine, G. L. Study of Strong Pressure Wave Propagation in a Two-Phase Bubbly Mixture, IMECE2010-40525, ASME 2010 International Mechanical Engineering Congress \& Exposition, Vancouver, British Columbia, November 10-12, 2010.
\end{enumerate}

\end{abstract}

\end{document}